\documentclass[amsmath,amssymb,aps,nofootinbib,showpacs,showkeys]{revtex4}
\usepackage{graphicx}
\usepackage{amsfonts}
\usepackage{bm}
\usepackage{amsmath}
\usepackage{amssymb}
\usepackage{color}
\usepackage[all]{xy}
\usepackage{verbatim}

\newcommand{\rmd}{\mathrm{d}} 
\newcommand{\ri}{\mathrm{i}} 
\newcommand{\tr}{\mathrm{tr}}
\newcommand{\ric}{\mathrm{Ric}}
\newcommand{\rt}{\mathrm{t}}
\begin{document}

\title{Extended Thermodynamics of Self-Gravitating Skyrmions}
\author{Daniel \surname{Flores-Alfonso}$^{1,}$}
\email[]{daniel.flores@correo.nucleares.unam.mx}
\author{Hernando \surname{Quevedo}$^{1,2,}$}
\email[]{quevedo@nucleares.unam.mx}
\affiliation{$^1$Instituto de Ciencias Nucleares,
Universidad Nacional Aut\'onoma de M\'exico,\\
AP 70543, Ciudad de M\'exico 04510, Mexico \\
$^2$Department of Theoretical and Nuclear Physics,
Kazakh National University, 
Almaty 050040, Kazakhstan}

\begin{abstract}
A modification of the hedgehog ansatz has recently led to novel exact black hole solutions
with selfgravitating SU(2) Skyrme fields. Considering a negative cosmological constant
the black holes are not asymptotically anti-de Sitter (AdS) but rather asymptote
to an AdS version of Barriola--Vilenkin spacetime.
We examine the thermodynamics of the system interpreting the cosmological constant
as a bulk pressure.
We use the standard counterterm method to obtain a finite Euclidean action.
For a given coupling of the matter action the system behaves as does a charged AdS black hole
in the fixed charged ensemble.
We find that in the limit case when the Skyrme model becomes a nonlinear sigma model
the system exhibits a first order phase transition of Hawking--Page type.
The universality class of these Einstein--Skyrme black holes is that of van der Waals.
\end{abstract}


\maketitle

\section{Introduction}

There have been recent developments in black hole thermodynamics by considering the cosmological constant as
a thermodynamic variable, which can be interpreted as a bulk pressure (for a recent review see \cite{chem}). 
The thermodynamic phase space is extended to include
pressure and volume into the fold.
In some cases, this thermodynamic volume can be interpreted as
a geometrical volume; however, this is not the case, in general \cite{cvj1}. 
Taking into account the cosmological 
constant in this way modifies the first law of thermodynamics. A relation was found by Smarr \cite{smarr} for asymptotically flat
black holes which corresponds to the Gibbs--Duhem equation. This procedure cannot be carried out in the same manner for
AdS black holes, say. Nonetheless, this is redeemed when the cosmological constant gains the status of a thermodynamic pressure.
In this extended setting, black hole thermodynamics is brought closer to the thermodynamics of ordinary substances
as it puts the first law in the same footing as the Smarr (or Gibbs--Duhem) relation. This has been shown to be valid in the case
of asymptotically AdS spaces, but it also holds for asymptotically Lifshitz spacetimes \cite{hyun}.
Moreover,  this approach also holds for spacetimes which are only locally asymptotically AdS \cite{cvj2}. 

In theories of semi-classical quantum gravity, where this approach has been applied, the gravitational mass of the
system corresponds to the thermodynamic enthalpy, rather than the energy. This means that in the limit
of vanishing cosmological constant the usual interpretation is recovered. In such a context,  the (Euclidean)
action serves as the fundamental equation for the thermodynamic system, i.e., all equations of state may be derived from it.
In general, gravitational actions diverge; notwithstanding, finite values can be obtained by using surface terms as counterterms
in the calculation \cite{surface}. The counterterm approach eliminates the need to carry out background subtractions, which is very
useful when ambiguities arise regarding the background that should be used to subtract. In some cases,
it is entirely unknown which background should be used. The approach taken in \cite{surface} is unique for (locally)
asymptotically AdS spaces as the couterterms depend crucially on the AdS curvature scale. 
The surface terms themselves
are universal and depend only on the dimension of spacetime and the cosmological constant.

Recently, black hole thermodynamics has been successfully applied to spacetimes which were previously unanticipated.
Some examples include C-metrics and Lorentzian Taub--NUT metrics\cite{Cmetrics,LTaubNUT}.
Within the Pleba\'{n}ski-Demia\'{n}ski family, which includes all Einstein--Maxwell black holes,
the acceleration and NUT parameters have been the most problematic. For example, they do not allow for
general type D metrics to be written in the usual Kerr--Schild form\cite{PS}. However, a recent approach
has allowed for the incorporation of these spacetimes into the standard thermodynamic treatment.
In accelerating black holes, a conical deficit is present, which can be interpreted as the consequence of
a cosmic string pulling on the black hole. In Taub--NUT spacetimes, the Misner strings can be thought as being the
reason behind conical deficits. The unifying concept for both spacetimes are conical defects. A fresh perspective
on the thermodynamics of spaces with conical defects has been proposed in reference \cite{condef}.

The early universe might have formed cosmic strings, which may be observed  due to their gravitational effects on the 
cosmic microwave background radiation or gravitational wave experiments. Cosmic strings are a certain type of topological defect.
They have been found to lead to interesting consequences when in the presence of primordial black holes \cite{coshole}.
In recent years, cosmic strings, domain walls and textures have received considerable attention.
These topological defects together with monopoles and skyrmions remain one of the most active fields in modern physics.
Topological defects arise in many elementary particle models and have found applications in cosmology \cite{vs}.
Skyrme was the first to devise a three dimensional topological defect solution arising from a nonlinear field theory \cite{skyrme}.
In the gravitational sector, the Einstein--Skyrme system has attracted considerable attention since
spherically symmetric black hole solutions with a nontrivial Skyrme field were found numerically \cite{dhs1}.
This marked the first counterexample to the black hole no-hair conjecture. It should be noted that this solution is stable against 
spherical linear perturbations \cite{dhs2}. 
In the present paper, we study the thermodynamic phase structure of a
black hole with Skyrme matter found by Canfora and Maeda \cite{cm}.
This solution possesses solid angle deficits very similar to the conical deficits due to cosmic strings.
The thermodynamics of black holes with conical defects has motivated us to study the thermodynamics
of spacetimes with solid angle deficits. 

This paper is organized as follows: In section \ref{ca}, we introduce the (AdS-) Barriola--Vilenkin spacetime which is a solution
to the Einstein field equations with nonlinear sigma model matter. We continue on to describe the dynamics of Einstein--Skyrme
systems and the Canfora--Maeda solution. In section \ref{thd}, we use the standard counterterm method to obtain the finite Euclidean
action for the system at hand and derive the relevant thermodynamic relations. We then provide a detailed description of the
phase structure of the Skyrme black hole. In section \ref{conc}, we close with a summary of our results, highlighting the novel aspects
of our work.

\section{Classical Aspects} 
\label{ca}

Many topological defects are represented by nonlinear sigma models, which are one of the most 
important nonlinear field theories. They have a vast application in physics which ranges from statical mechanics
to gravitation, especially string theory. Some examples are Nambu--Goldstone bosons, the superfluid $^3$He and the quantum Hall effect.
The Barriola--Vilenkin spacetime was built to support global monopoles \cite{bv}; they are an example of nonlinear sigma model
matter harbored in Einstein backgrounds. A close configuration is the Gibbons--Ruiz Ruiz black hole \cite{global}, 
which shares its asymptote with that of Barriola--Vilenkin.

Scaling arguments stemming from Derrick's renowned theorem show that nonlinear sigma models do not admit static soliton solutions in 3+1 dimensions.
Skyrme constructed his model exactly to circumvent this result and did so by adding higher derivative terms to the action.
This makes manifest that skyrmions behind horizons share asymptotic behavior with their corresponding sigma model limit.
It should be noted that skyrmions describe an entirely different class of topological defects. 
Skyrmions and other ``classical lumps'' with horizons have been investigated in the literature \cite{lumps,hhadt}; however,
their role in the AdS context is less clear \cite{deng}. 

Before continuing any further, we shortly review spacetimes that are asymptotically AdS with a solid angle deficit in comparison to AdS itself. 
We write the AdS generalization of the Barriola and Vilenkin global monopole spacetime as
\begin{equation}
 \rmd s^2=-B(r')\rmd t'^2+A(r')\rmd r'^2+r'^2(\rmd\theta^2+\sin\theta^2\rmd\phi^2),\label{generic}
\end{equation}
with the metric functions given by
\begin{equation}
 B(r')=A(r')^{-1}=1-\alpha+\frac{r'^2}{l^2}.
\end{equation}
Hereafter, we refer to this spacetime as AdS Barriola--Vilenkin and abbreviate it as AdS-BV.
Here $l$ is the AdS curvature radius and is related to the cosmological constant by $\Lambda=-3/l^{2}$.
Now, making a coordinate change
\begin{subequations}
 \begin{align}
 r'&=r(1-\alpha)^{1/2},\\
 t'&=t(1-\alpha)^{-1/2},
 \end{align}
\end{subequations}
the above line element becomes
\begin{equation}
 \rmd s^2=-\left(1+\frac{r^2}{l^2}\right)\rmd t^2+\left(1+\frac{r^2}{l^2}\right)^{-1}\rmd r^2+r^2(1-\alpha)(\rmd\theta^2+\sin\theta^2\rmd\phi^2).
\end{equation}
Note that for vanishing $\alpha$ we recover the AdS geometry. 
For constant $t$ and $r$ the above metric describes a sphere with solid angle deficit of $4\pi\alpha$.
The Gibbons--Ruiz Ruiz black hole has a line element similar to (\ref{generic}), but with metric functions given by
\begin{equation}
 B(r')=A(r')^{-1}=1-\alpha-\frac{2m'}{r'}.
\end{equation}

The class of asymptotically flat spacetimes with an angle deficit $\alpha$ has been studied by Nucamendi and Sudarsky \cite{ns}.
Therein, they find an Arnowitt--Deser--Misner (ADM) mass generalization given by $m=m'(1-\alpha)^{-3/2}$.
In this paper, we focus on a recent black hole found by Canfora and Maeda \cite{cm}, which has Skyrme matter and includes a cosmological
constant. Their metric has this same type of solid angle deficit and specializes to the Gibbons--Ruiz Ruiz spacetime.

\subsection{Einstein--Skyrme Systems}
\label{es}

In this work, we concentrate on the thermodynamic behavior of a black hole with nonlinear scalar matter.
The system is a four dimensional Einstein--Skyrme configuration and includes a cosmological constant. 
The matter content are scalar fields which fulfill a generalized hedgehog ansatz.
This section is dedicated to describing the SU(2) Skyrme model; for more details, see \cite{heuslerbook}.

We write the basic action as
\begin{equation}
I[g,U]=-\frac{1}{16\pi G}\int d^4x\sqrt{-g}(R-2\Lambda)-\frac{K}{2}\int \tr\left(\frac{1}{2}A\wedge\star A+\frac{\lambda}{8}F\wedge\star F\right), 
\label{esk}
\end{equation}
where the fundamental field $U$ is an SU(2)$-$valued scalar field and $A$ is the pullback of the Maurer-Cartan form of SU(2) by $U$.
It follows that $A= U^{-1}\rmd U$ is an ${\mathfrak{su}}(2)-$valued one$-$form and complies with $A\wedge A=-\rmd A$. It is also useful to define $F=A\wedge A$.
The first term in the matter action of (\ref{esk}) is that of a nonlinear sigma model. Notice that it is a quadratic term and so the
second term can be recognized as a quartic term or as a higher curvature term, whichever is more convenient. We emphasize that the second 
term, proportional to $\lambda$, was added by Skyrme to deform the nonlinear sigma model action.
Notice that $A$ is reminiscent of a Yang--Mills pure gauge potential and that the equation $F=-\rmd A$
might lead to similarities with Maxwell matter. We also comment that in a pion model ($\lambda=0$) $K$ is related to the pion decay constant
and since pions are pseudo-Goldstone bosons, then $K$ characterizes the symmetry breaking present in the system. In a global monopole,
$K$ is related to the solid angle deficit $\alpha$ in spacetime --- as compared to a Minkowski background.

It is standard to parametrize the ${\mathfrak{su}}(2)$ directions using Pauli matrices $\sigma^i$ or alternatively by $\rt^i=-\ri\sigma^i$.
So the ${\mathfrak{su}}(2)-$valued one$-$form $A$ may be decomposed as $A=A^{i}\otimes \rt^{i}$ or as $A=A_{\mu}\otimes dx^{\mu}$.
The $\rt^i$ can also be used together with unity $\bf{1}$ as a base for $\mathbb{R}^4$ and so a point $y$ on SU(2)$\approx S^3$
can be parametrized by
\begin{equation}
 y=Y_0{\bf{1}}+Y_i\rt^i, \label{s3}
\end{equation}
as long as $Y_0^2+Y_1^2+Y_2^2+Y_3^2=1$.

Turning to the action once more, the equations of motion are
\begin{equation}
 \ric(g)-\frac{R}{2}g+\Lambda g=8\pi G T,
\end{equation}
where the energy-momentum tensor $T$ has components
\begin{equation}
T_{\mu\nu}=-\frac{K}{2}\tr\left[\left(A_{\mu}A_{\nu}-\frac{1}{2}g_{\mu\nu}A_{\alpha}A^{\alpha}\right)
 +\frac{\lambda}{4}\left(F_{\mu\alpha}F_{\nu}^{~\alpha}-\frac{1}{4}g_{\mu\nu}F_{\alpha\beta}F^{\alpha\beta}\right)\right],
\end{equation}
and
\begin{equation}
 \rmd\star A-\frac{\lambda}{4}[A,d\star F]=0.
\end{equation}
Notice that sigma model solutions, which satisfy $\rmd\star A=0$, can be promoted to Skyrme model solutions provided they comply additionally with $[A,d\star F]=0$
so that the above equation is satisfied.

Following reference \cite{cm}, we also define a symmetric tensor ${\cal S}=-1/2~\tr(A\otimes A)$ with  components
\begin{equation}
 {\cal S}_{\mu\nu}=\delta_{ij}A^i_{\mu}A^j_{\nu}.\label{s}
\end{equation}
So the energy-momentum tensor is now written as
\begin{equation}
T_{\mu\nu}=K\left[\left({\cal S}_{\mu\nu}-\frac{1}{2}g_{\mu\nu}S\right)
 +\lambda\left(S{\cal S}_{\mu\nu}-{\cal S}_{\mu\alpha}{\cal S}_{\nu}^{~\alpha}-\frac{1}{4}g_{\mu\nu}(S^2-{\cal S}_{\alpha\beta}{\cal S}^{\alpha\beta})\right)\right],
\end{equation}
where $S$ is the trace of ${\cal S}$. This last depiction explicitly breaks the similarities to Yang--Mills matter
and makes the degree of nonlinearity in Skyrme matter manifest. Although, as pointed out in \cite{cm}, the
Skyrme contribution is traceless very much like Maxwell matter.
The matter equations are now portrayed as
\begin{equation}
 \nabla^{\mu}\left[A_{\mu}+\lambda\left(SA_{\mu}-{\cal S}^{\nu}_{~\mu}A_{\nu}\right)\right]=0.
\end{equation}
The nonlinearity inherit to these equations makes finding exact solutions difficult. However, the use of simplifying Ans\"atze
makes the equations more tractable.

\subsection{The Canfora--Maeda Solution}

The exact solution under discussion was built by generalizing the hedgehog Ansatz in such a way that the fields themselves
need not reflect spherical symmetry, but their energy-momentum tensor does. This ultimately makes the metric resemble
that of the Reissner--Nordstr\"om geometry. The fundamental scalar field $U$ is given by the map
\begin{equation}
U:(t,r,\theta,\varphi)\mapsto \rt^r=\cos\theta \rt^3+\sin\theta\sin\varphi \rt^2+\sin\theta\cos\varphi \rt^1. \label{U}
\end{equation}
Notice that $\rt^r$ has Frobenius norm -2 as all the other $\rt^i$; this justifies our notation. Notice that the $\rt^i$
 are used here as in Eq.(\ref{s3}) so that $\rt^r$ represents the unit radial vector in $\mathbb{R}^4$;
in other words, it is in correspondence with the positions of $S^3$.  

The geometry generalizes the Schwarzschild-AdS spacetime, which is recovered by setting $K=0$, and is given by
\begin{subequations}
\begin{align}
 \rmd s^2&=-f(r)\rmd t^2+f(r)^{-1}\rmd r^2+(1-\alpha)r^2(\rmd\theta^2+\sin^2\theta \rmd\phi^2),\label{metric}\\
 f(r)&=1-\frac{2Gm}{r}+\frac{q^2}{r^2}+\frac{r^2}{l^2}.\label{f}
\end{align} 
\end{subequations}
However, $q$ is not an integration constant as in the Reissner--Nordstr\"om black hole, but is fixed by the
coupling constants of the theory. Here $\alpha$ parametrizes the solid angle deficit $4\pi\alpha$ and is given by $\alpha=8\pi GK$.
Moreover, $q^2$ is just shorthand for $\alpha\lambda/2(1-\alpha)^2$ and $m$ is the
Nucamendi--Sudarsky mass. A quasi-local calculation \cite{abbott,deser,tekin,KimKulkarniYi,GimKimYi} yields the Abbott--Deser--Tekin (ADT) mass as
\begin{equation}
 M=m(1-\alpha).
\end{equation}
For asymptotically AdS spacetimes, the ADT mass is calculated  as the ADM mass, except that the background metric is not flat
and the lapse is not unity. It also yields the standard result for the energy per unit length of a cosmic string, which
is proportional to the angle deficit\cite{hhadt}. The exterior of a cosmic string is Minkowski, but with a solid angle deficit.

\section{Thermodynamics} \label{thd}

As mentioned in the previous section, the Canfora--Maeda black hole generalizes the AdS-Schwarzschild spacetime. 
This itself
motivates investigating its thermodynamics. As Hawking and Page have shown, there are interesting
phase transitions in the AdS-Schwarzschild system \cite{hpads}.
In our thermodynamic treatment, we shall consider the cosmological constant as a thermodynamic variable.
The matter content is scalar and so no modifications from the matter sector are expected for the
first law of thermodynamics. However, the Skyrme matter does contribute to the Smarr relation \cite{hs}
and in an AdS context, considering the cosmological constant as a thermodynamic pressure, gives the generalized relation  \cite{hyun}.

\subsection{Action and Counterterms}

We use the path integral approach to semi-classical quantum gravity and consider the cosmological constant as a thermodynamic
variable. We study the analytic continuation ($t\to\ri\tau$) of the black hole solution (\ref{metric}) and identify
the period of the imaginary time $\beta$ with the inverse temperature. The period is fixed by the amount that makes the
Euclidean solution regular. Euclidean solutions are required to be regular everywhere so that they serve as a saddle point
to approximate the path integral. Since the black hole horizon is transmuted into a bolt when the analytic continuation is carried out,
then the root of the metric function (\ref{f}) $r_+$ will lead in general to a conical singularity. As mentioned above, this is resolved by fixing the period
of the Euclidean time circle by
\begin{equation}\label{betatemp}
 \beta=\frac{4\pi}{f'(r_+)}.
\end{equation}
Let us recall that upon analytical continuation one must check first for curvature singularities.
For the solution at hand the Kretschmann scalar is infinite at $r=0$.
Any Euclidean sheet which is expected to dominate the gravitational path integral cannot contain the region $r=0$. Moreover, on the Lorentzian sheet
null surfaces such as the horizon may exist without problem. However, when the signature is positive definite a place where the metric function vanishes
is a degenerate region. The only way for a positive definite manifold to have a region where, say $f(r_+)=0$, is that it resembles a cone, or cigar.
Since the direction which degenerates at the tip of this cone is the Euclidean time, then it must be this direction which is periodic. This justifies
Eq.(\ref{betatemp}), regularizes the horizon and avoids the curvature singularity. Although there is an angular deficit in spacetime 
which extends to infinity, the only place where its effect
is locally observable is at the curvature singularity which (\ref{betatemp}) avoids.

The Skyrme map (\ref{U}) is unaffected by the analytic continuation and we calculate the Skyrme contribution to the energy of the system as
\begin{equation}
 E=-\frac{K\lambda}{16\beta}\int\limits_{\cal M} \rmd^4x\sqrt{g}~\tr\left\langle F,F\right\rangle=\frac{2\pi K\lambda}{r_+(1-\alpha)}=\frac{q^2(1-\alpha)}{2Gr_+}. \label{skyrmE}
\end{equation}
This expression is like the electric contribution to the energy in a charged black hole with a global monopole \cite{deng}.
In such a black hole, we would have an electric potential difference of $\Phi=q/2r_+$ and a total electric charge of $Q=q(1-\alpha)/G$.
This implies that the Maxwell sector provides and amount of energy given by $\Phi Q=q^2(1-\alpha)/2Gr_+$.
The analogy we are drawing here makes sense because $q^2$ is the amount that appears multiplying the $1/r^2$ term in the metric
function (\ref{f}).
Furthermore, the Skyrme term in the action needs no renormalization, as is the case of Maxwell matter in four dimensions.

Before continuing with the action calculation, we recall that in the \emph{extended} thermodynamics
we are considering the cosmological constant as a canonical variable and so the effective pressure is given by
\begin{equation}
 P=-\frac{\Lambda}{8\pi G}=\frac{3}{8\pi G l^2}.
\end{equation}
So the appearance of the cosmological scale in the finite action will point to quantities involving the bulk pressure.

When it comes to black hole thermodynamics, the gravitational path integral in the saddle point approximation has
a long history in the quantum gravity literature. Solutions related to AdS were studied in this context in \cite{hpads,hhadt} about the same time, 
but in parallel to the use of the cosmological constant as a pressure \cite{henneaux1,teitelboim,henneaux2}.
The persistent problem with this approach is that typically the Euclidean action diverges. 
The counterterm method we focus on is standard practice\cite{surface}
and it allows to solve this problem. In four dimensions, a finite gravitational action is achieved through
\begin{eqnarray}
 I^{\rm{ren.}}_G=-\frac{1}{16\pi G}\int\limits_{\cal M} \rmd^4x\sqrt{g}\left(R+\frac{6}{l^2}\right)
 -\frac{1}{8\pi G}\int\limits_{\partial {\cal M}} d^3x\sqrt{h} {\cal K}\nonumber\\
 +\frac{1}{8\pi G}\int\limits_{\partial {\cal M}} d^3x\sqrt{h} \left[\frac{2}{l}+\frac{l}{2}{\cal R}\right].
 \label{renaction}
\end{eqnarray}
Here the geometry of $g$ on ${\cal M}$ induces a metric $h$ on the boundary $\partial {\cal M}$.
In this approach, the boundary at infinity is held fixed to obtain the Einstein field equations. The action is not only the usual
Einstein--Hilbert action, but it also contains the Gibbons--Hawking surface term, easily recognized in the above equation by the trace of the extrinsic curvature ${\cal K}$ of the boundary as embedded in ${\cal M}$. The surface terms at the end of the renormalized action are counterterms which
are unique in the AdS context, as they depend on the AdS scale. These terms are sufficient in four dimensions, but in higher dimensions
extra terms are required, which are explicitly known at least up to dimensions relevant in the AdS context.  
Notice that the counterterms
are covariant and depend only on, e.g., the Ricci scalar ${\cal R}$ of the boundary. Moreover, it has been shown that
this regularization is equivalent to the addition of a Gauss--Bonnet term\cite{miskovic-olea}.

For the Canfora--Maeda black hole (\ref{metric}-\ref{f}), we get the following finite gravitational action
\begin{equation}
 I^{\rm{ren.}}_G=\frac{\beta}{2G}\left(Gm(1-\alpha)+r_+\alpha-\frac{r_+^3(1-\alpha)}{l^2}\right)
\end{equation}
On the other hand, the quadratic part of the matter action diverges. This is the part we have called the sigma model
segment of the scalar action. So, an additional counterterm must be added to renormalize this divergence.
Scalar matter and global monopole counterterms have been studied before in \cite{kostas,radu}.
However, the counterterm we use for the matter content is
\begin{equation}
 I^{\rm{ct}}_M=\frac{K}{2}\int\limits_{\partial {\cal M}} d^3x \sqrt{h}\left[\frac{l}{2}\tr\langle A, A\rangle\right]
 =-K\int\limits_{\partial {\cal M}} d^3x \sqrt{h} \left[\frac{l}{2}S\right]. \label{skyct}
\end{equation}
This counterterm is of kinetic type and is reminiscent of the one found for scalars in Lifshitz spacetimes \cite{andrade}.
Since $A$ does not depend on $r$, there is no ambiguity in the above equation. 
Notice that written in this way, the counterterm, as written in the left hand side, shares a similar structure as
 the ones in \cite{marika}. Observing the right hand side, we notice that it has the same structure as the very last term in 
(\ref{renaction}).
Finally, we write the action of the Einstein--Skyrme system as
\begin{equation}
 I^{\rm{ren.}}=\frac{\beta}{2G}\left(Gm(1-\alpha)-\frac{r_+^3(1-\alpha)}{l^2}+\frac{q^2(1-\alpha)}{r_+}\right). \label{eaction}
\end{equation}
We now portray the imaginary time period as
\begin{equation}
 \beta=\frac{2\pi r_+^2(1-\alpha)}{Gm(1-\alpha)-q^2(1-\alpha)/r_++r_+^3(1-\alpha)/l^2}, \label{beta}
\end{equation}
which means that we can rewrite the action in the following form
\begin{equation}
 I=\beta m(1-\alpha)-\frac{\pi r_+^2(1-\alpha)}{G}.
\end{equation}
Immediately, we compute the state variables of the system, as the Gibbs free energy is $I/\beta=H-TS$
\begin{subequations}
 \begin{align}
  H&=\left(\frac{\partial I}{\partial \beta}\right)_P=m(1-\alpha)=M, \label{enthalpy}\\
 S&=\beta\left(\frac{\partial I}{\partial \beta}\right)_P-I=\frac{\pi r_+^2(1-\alpha)}{G}=\frac{A_h}{4G},\quad\rm{and} \label{entropy}\\
 V&=\frac{1}{\beta}\left(\frac{\partial I}{\partial p}\right)_\beta=\frac{4\pi r_+^3(1-\alpha)}{3}. \label{volume}
 \end{align}
\end{subequations}
The enthalpy $H$, entropy $S$ and thermodynamic volume $V$ indeed fulfill the first law of thermodynamics
\begin{equation}
 \rmd H=T\rmd S+V\rmd P.
\end{equation}
Moreover, the expression (\ref{beta}) takes on the new form
\begin{equation}
 \frac{H}{2}=TS-PV+E ,
\end{equation}
which is recognizable as the Smarr relation and generalizes equation (97) in \cite{hs} for vanishing angular momentum.
Certainly, the path integral approach is consistent with the mass variation approach to thermodynamics and the quasi-local 
formalism \cite{kastor,hs,hyun}. 
We recollect that in the perfect fluid interpretation of the cosmological constant, the energy density is 
$\rho=-P=\Lambda/8\pi G$. Removing a portion of spacetime to form a black hole of volume $V$
thermodynamically costs an amount $PV$. This formation energy is naturally captured in the enthalpy
and, from the gravitational point of view, this is reflected in the black hole mass \cite{kastor,cvj1}.

To sustain global monopoles and skyrmions such as (\ref{U}), we must cut out a region of spacetime given the angle deficit $4\pi\alpha$.
So, to form a black hole in AdS with an angle deficit, a smaller volume needs to be removed. 
The difference in volumes is, of course, given by the cone over the deficit area $1/3(4\pi\alpha r_+^2)r_+$.
The solid angle deficit in spacetime affects all extensive thermodynamic quantities. If we recall the holographic
principle, this is exactly what is expected. The entropy is given by the horizon area and so
the proportionality $(1-\alpha)$ in (\ref{entropy}) and (\ref{volume}) becomes clear.
Since the mass and Skyrme energy can be written in terms of surface integrals \cite{hs}, they too 
will have this same $(1-\alpha)$ factor; see (\ref{enthalpy}) and (\ref{skyrmE}). As we mentioned above,
Skyrme energy is comparable to electric energy. The extensive variable for Maxwell configurations is the
electric charge which, in general, may be written as a surface term. This is observed in \cite{deng}, where
the electric charge possesses the same proportionality factor as all other global charges.

\subsection{Black Hole Chemistry}

The inclusion of pressure, volume and enthalpy into black hole thermodynamics has led to a different understanding
of gravitational systems. A prevalence of van der Waals universality classes can be found in the literature and
black hole mechanics is comparable to chemical systems.

\begin{figure}[ht]
\includegraphics[scale=0.45]{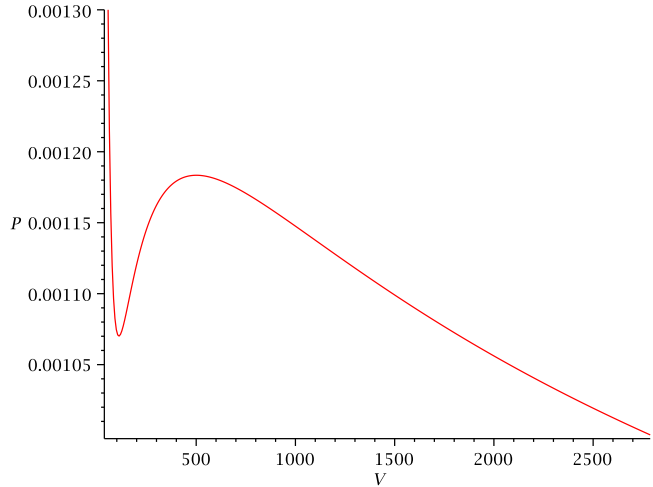}
\caption{The bulk pressure of the Skyrme black hole system is displayed as a function of the thermodynamic volume.
The plot is in accordance with the critical behavior of a van der Waals fluid. In this figure we have chosen $T=0.05$
and $q^2=2.25$.
\label{PVvdW}}
\end{figure}

\begin{figure}[ht]
\includegraphics[scale=0.45]{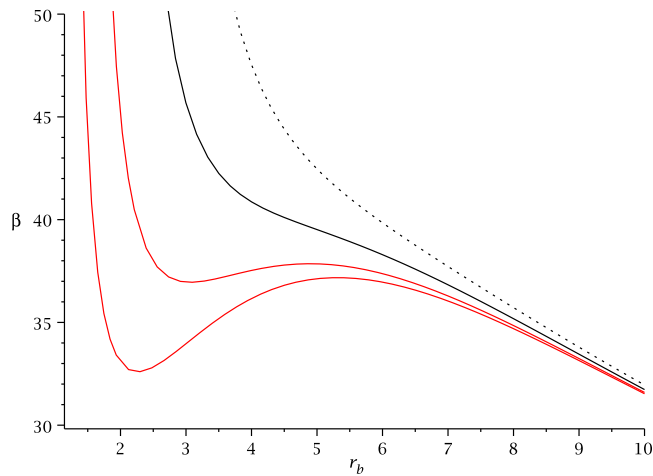}
\caption{The curves displayed above are seen to be analogous to a $P-V$ diagram
of van der Waals fluids. The inverse temperature is plotted versus the bolt radius for
different values of the Skyrme coupling constant. Here we have chosen $G=2\alpha=1$ and $l=10$.
The black curve represents the critical value of $\lambda$, the dotted curve is supercritical
and the red curves are subcritical.\label{betavsr}}
\end{figure}

Starting from equation (\ref{beta}) and recalling that $f(r_b)=0$, we may write an expression for the pressure as a function of its volume
\begin{equation}
 P=\frac{T}{2r_b}-\frac{1}{8\pi r_b^2}+\frac{\alpha\lambda}{16(1-\alpha)^2\pi r_b^4},
\end{equation}
given that
\begin{equation}
 r_b(V)=\left(\frac{3V}{4\pi}\right)^{1/3}.
\end{equation}
The Einstein--Skyrme system exhibits the famous van der Waals $P-V$ curve in Figure \ref{PVvdW}.
In charged AdS black holes, a critical charge can be found where the temperature's turning points appear or disappear.
In the Canfora--Maeda solution (\ref{U}-\ref{f}) the equivalent is a critical value of the Skyrme coupling constant. 
Figure \ref{betavsr} portrays how the system conducts itself for different couplings.

\subsubsection{Hawking--Page transition}

\begin{figure}[ht]
\includegraphics[scale=0.35]{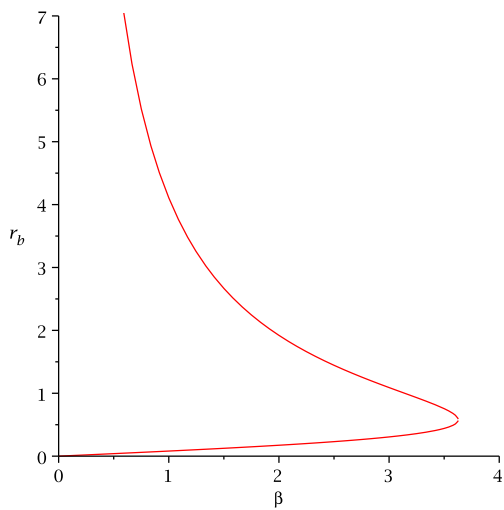}
\caption{This figure portrays the bolt radii of the Euclidean AdS black holes versus the inverse temperature; here, we have used $l=1$.
The union of both branches is shown. They connect at the temperature $T=\sqrt{3}/2\pi l$. \label{rbeta0}}
\end{figure}

In the special limit $\lambda=0$,
 the matter field reduces to an $SU(2)$ nonlinear sigma model.
The equation of motion for the scalar matter is now
\begin{equation}
 d\star A=0 \ .
\end{equation}
The fundamental $SU(2)-$valued scalar field $U$ is still given by equation (\ref{U}) and satisfies the above
equation of motion. The metric, in this limit, is the AdS version of Gibbons and Ruiz-Ruiz
\cite{global} and corresponds
to AdS-Schwarzschild with a solid angle deficit. 
Its Euclidean counterpart has a bolt radius
given by the following quadratic equation
\begin{equation}
 3\beta r_b^2-4\pi l^2r_b+l^2\beta=0, \label{quadratic}
\end{equation}
which means that given a temperature and a pressure there are two branch solutions
\begin{equation}
 r_{b\pm}=\frac{2\pi l^2}{3\beta}\left(1\pm\sqrt{1-\frac{3\beta^2}{4\pi^2l^2}}\right).
\end{equation}
We therefore speak of a large $(r_{b+})$ and small $(r_{b-})$ black hole configuration.
These branch solutions are plotted in Figure \ref{rbeta0}.
Moreover, there is a minimum available
temperature for the black holes to exist, the size of the branch black holes coincide at this temperature, which is given by
\begin{equation}
 \beta_{\rm{max}}=\frac{2\pi l}{\sqrt{3}}.
\end{equation}
These black holes are precisely the AdS Schwarzschild black holes \cite{hpads} when we set 
$\alpha=0$. Further, still
the AdS-BV spacetime reduces to AdS in this special case. From the AdS perspective, there is a region of
spacetime that has been cut away. In other words, this is only appreciated when compared to the AdS space.
Even when this point of view is taken, the black hole behavior remains intact ---  despite there being a ``removed region''. 
To further demonstrate this, we explore the phase structure of the system.

\begin{figure}[ht]
\includegraphics[scale=0.35]{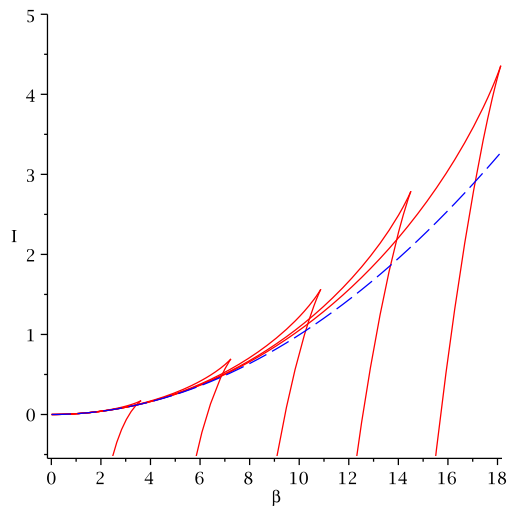}
\caption{The Euclidean action, which is proportional to the Gibbs free energy, for different values of pressure $P=3/8\pi G l^2$.
We have chosen the values $l=1,2,3,4,5$ with $G=2\alpha=1$. The solid lines from left to right
decrease in pressure while the dashed curve is the zero pressure limit. \label{action0}}
\end{figure}

Figure \ref{action0} shows the Euclidean action of the black hole branches versus the inverse temperature for various values of the pressure.
Notice that a zero temperature a solution is only possible when $m=0$; this case corresponds to AdS-BV for which
the Euclidean action vanishes. Comparing the black hole action values, we see that the large black hole
is always preferred over the small one. This indicates that the small black hole is an unstable configuration of the system.
There is also a set of temperatures for which the Euclidean action is negative, meaning that it is lower than that of
AdS-BV. This is to say that for high temperatures the preferred phase of the system is a large black hole
while for low temperatures the AdS-BV phase dominates.

We point out that at every fixed pressure, while transitioning between AdS-BV and the large black hole, there is a discontinuity in the entropy.
This labels the phase transition as first order and parallels the Hawking--Page transition in AdS-Schwarzschild\cite{hpads}
and between Taub--NUT and Taub-Bolt spaces found in \cite{cvj2}. For vanishing cosmological constant our results are consistent with those obtained in the 
thermodynamics of global monopole systems \cite{jyw,yu}.

\subsubsection{Skyrme AdS black holes}

After examining the sigma model limit of the system, we turn back to the investigation of
 the Skyrme coupling constant $\lambda$.
For fixed values of $\alpha$, the $q^2$ constant which enters the metric (\ref{metric}) is tuned by the 
Skyrme parameter $\lambda$. The geometry is of Reissner--Nordstr\"om type with the exception
that $q^2$ is not an integration constant, but is related to the coupling constant.
The Canfora--Maeda solution is like the charged AdS black holes investigated in 
\cite{swallowtail,kubizmann}.
However, a chief difference is the manipulation of the $q^2$ parameter. For charged black holes, a fixed potential ensemble
has been studied as well as a fixed charge ensemble. The behavior of the present system is comparable to AdS black holes
with a fixed charge.

The bolt radii of the black hole must comply with the following quartic equation
\begin{equation}
 3\beta r_b^4-4\pi l^2r^3+l^2\beta r^2-l^2q^2\beta=0.
\end{equation}
Notice that when $q^2=0$, we recover equation (\ref{quadratic}). 
In this limit, two of the four solutions
of the quartic equation become repeated and are null. 
It turns out that one of the four solutions is in general negative and so
unphysical; this is one of the solutions which is nullified together with $q^2$. The other solution is physical
and in that limit vanishes and represents AdS-BV. 
Algebraically, all this can be deduced from the discriminant
of the quartic equation. Moreover, only when the discriminant is positive, we will have four real roots which represent three physical
solutions: a small, large and intermediate black hole.

\begin{figure}[ht]
\includegraphics[scale=0.45]{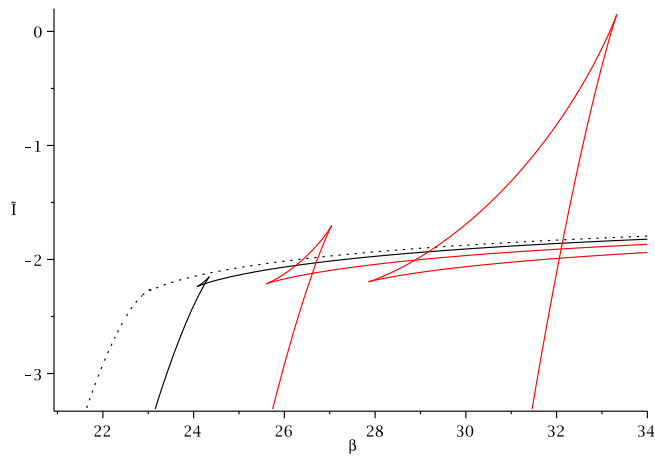}
\caption{The Euclidean action of the Einstein--Skyrme system as a function of the inverse temperature
for $G=2\alpha=q^2=1$ and different values of the pressure $P=3/8\pi G l^2$. The black lines correspond to $l=6$ (dotted) and $l=6.4$ (solid)
while the red lines are for $l=7.2, 9$.
\label{IminusIe}}
\end{figure}

For $\beta\to\infty$ one of the roots approaches
\begin{equation}
 r_e=\frac{1}{6}\sqrt{-6l^2+6l\sqrt{l^2+12q^2}}.
\end{equation}
Hawking radiation will be absent for a black hole of this size.
This represents a zero temperature black hole with a single degenerate horizon. This is the lowest entropy
configuration of the system and corresponds to an extremal black hole such that
\begin{equation}
 \lim_{\beta\to\infty}\left(\frac{I}{\beta}\right)=M_e,
\end{equation}
meaning that the free energy is given by the extremal black hole's mass.
Notice that taking $q^2=0$ yields $r_e=0$ and so the extremal black hole turns into AdS-BV in the nonlinear sigma model limit.
Finite temperature black holes are an excitation of this extremal configurations.
Thus, we consider an action $\tilde{I}=I-I_e=\beta(M-M_e)-S$ which yields thermodynamic equations of state
\begin{subequations}
 \begin{align}
 H&=\left(\frac{\partial \tilde{I}}{\partial \beta}\right)_P=M-M_e,\\
 S&=\beta\left(\frac{\partial \tilde{I}}{\partial \beta}\right)_P-\tilde{I}=\frac{A_h}{4G},\quad\rm{and}\\
 V-V_e&=\frac{1}{\beta}\left(\frac{\partial \tilde{I}}{\partial p}\right)_\beta=\frac{4\pi}{3}(r_b^3-r_e^3). \label{tilde-thd}
\end{align}
\end{subequations}
The thermodynamic variables fulfill the first law in the form $\rmd H=T\rmd S+(V-V_e)\rmd P$. 
Using the extremal black hole as a starting point and cutting out spherical regions, 
 the horizon forms
a finite temperature black hole where the energy of formation is given by $P(V-V_e)$.
In figure \ref{IminusIe}, the action difference $I-I_e$ is plotted an shows a distinctive swallowtail behavior
indicating a region where phase transitions occur. Paralleling the $\lambda=0$ behavior, we can read off from figure \ref{IminusIe}
that the intermediate black hole branch always has higher free energy than any of the other branches. This phase is never
statistically preferred by the system, it is unstable. For a fixed pressure, the large and small black hole phases have coinciding free energies at 
a single temperature, the coexistence temperature.

Figure \ref{IminusIe}, which has $q=1$, is generic in the sense that higher values of $q$ pull the graph toward more negative values 
when the swallowtail is present.
Figure \ref{betavsr} encapsulates the information that there is a critical range where the swallowtails appear.
As $q$ gets smaller but remains positive, the swallowtail doe rise toward positive values remaining on the lower half of the plane.
As expected, the more one approaches zero the more the swallowtail will morph into the Hawking--Page characteristic curve. Up until now we have maintained $q$ positive
since it couples a (quartic) kinetic term to the rest of the action. To avoid instabilities and ghosts it must remain so. However, thinking of an analogue 
of the Mexican hat potential, we explore negative values of $q$.
This sector shows a completely different dynamics than the one described up until now. A small and a large black hole are present, but no intermediate black hole and hence no van der Waals
transition. Since AdS-BV is also not an available phase, there is also no Hawking--Page transition. One of the phases is always favored thermodynamically and so the sector
is devoid of phase transitions.

\begin{figure}[ht]
\includegraphics[scale=0.35]{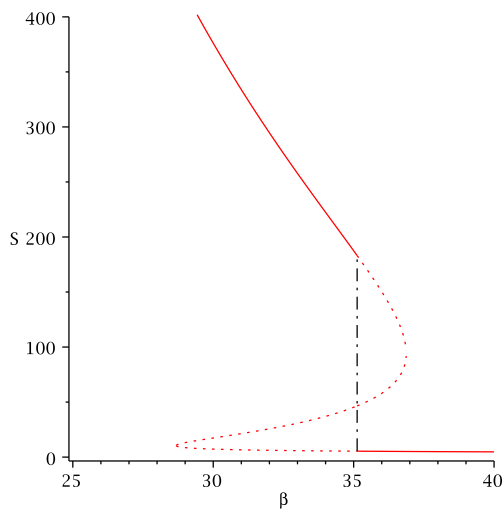}
\caption{The entropy of the system in terms of the inverse temperature. The multiple black hole branches
are represented in the plot. The curve is solid when the entropy corresponds to a phase statistically preferred by the system ---
when the phase has lower free energy than any other phase. The dashed curve is the complement and so the entropy does not
follow this curve. The discontinuity in the entropy is marked with a black dot-dashed curve.
\label{Svsbeta}}
\end{figure}

Consider a nearly extremal (small) black hole and raise its temperature  until  the coexistence temperature is reached.
Raising the temperature even further will cause the black hole to transit into a large black hole. The size of the black hole
blows up discontinuously and as a consequence so does its entropy. 
The thermodynamic system exhibits a first order phase transition
at the coexistence temperature. The entropy's behavior is presented in figure \ref{Svsbeta}.

\section{Conclusions}
\label{conc}

In this paper we investigate the extended thermodynamic behavior of a self-gravitating skyrmion in Einstein--Skryme theory. 
The spacetime is interpreted as a black hole which has swallowed a skyrmion, in a special limit the system becomes
a black hole with a global monopole in its interior. The approach we use is standard. We use the counterterm method to obtain
the Euclidean action --- which in turn yields the free energy of the system. However, there is a new aspect to the renormalization
performed here. The Skyrme sector is only regularized with the term in equation (\ref{skyct}), which we did not find in the existing literature.
The term is of kinetic type since the Skyrme action is also of this type. Common counterterms for scalar fields cannot be used here
because of the model's non-scaling properties.

The thermodynamics of the Canfora--Maeda solution we investigate has not been fully explored previously. The Euclidean approach yields
an enthalpy for the Einstein--Skyrme system. We use the covariant quasi-local ADT method to obtain a value for the systems mass and find it
to coincide with the enthalpy of the system. This further supports the idea that the cosmological constant enters thermodynamics through
the formation energy of the system. This is contrary to the approach taken in references \cite{Cmetrics,condef,LTaubNUT},
where related thermodynamics are carried out. There, the conformal completion method is used to obtain the mass. For the sake of comparison
we have calculated the mass through this method and find it to coincide with the ADT energy. In these works the thermodynamics of various
spaces with conical defects are carried out. It is in this way they relate to spaces with solid angle defects such as the ones present here.
The present Skyrme system has a background metric which is of Reissner--Nordstr\"om type with the exception that the factor in the $r^{-2}$ term
in the metric function is not an integration constant. It does not represent any electromagnetic charge and is fixed by the coupling constants 
of the theory. This contributes to the similarity between the dynamics of the Skyrme system and charged black holes. The areal deficit in the spacetime
is comparable to spacetimes which possess cosmic strings. There the string tension is directly related to the conical deficit. In the references mentioned
just above a first law of thermodynamics is posed where the string tension varies and gives rise to a new thermodynamic potential.
Here the solid angle deficit is fixed by the coupling constants and so this procedure cannot be carried out.

The thermodynamics of nonlinear matter have been studied before for example in Einstein--Skyrme systems but also recently
in other types of nonlinear scalar models. The latter in an extended setting considering thermodynamic volume and the cosmological
constant as a bulk pressure. To our knowledge this is the first time an extended Smarr relation has been derived for
spacetimes containing Skyrme matter. We generalize previous results and find that the thermodynamic volume allows for a geometric interpretation.
In other words, the asymptotically locally AdS solution we consider can be thought of as formed from a ground state spacetime by removing 
a portion of its volume. This encapsulates forming a black hole where there was none, making it larger and also removing a wedge from it entirely
--- which corresponds to the solid angle deficit. We have extended the original sense of the formation energy concept \cite{kastor} to apply
for ``missing'' wedges coming from topological defects.

We also mention that charged monopole black holes have been studied recently in the literature and their behavior has been found to be of
van der Waals type. This further supports the idea that the black hole behavior within the setting of extended thermodynamics is universal.
Although very similar to charged black holes lacking scalar matter some central differences may found such as the location of critical points.
It has been found that these points depend on the exact value of the symmetry breaking parameter, $K$ in our notation. Nonetheless, aspects such as
the law of corresponding states are blind to this parameter. Comparing this to our current work, we find the result to hold as well. 
At this point we stress that the Skyrme system at hand is electromagnetically neutral yet the nonlinearity of the scalar matter yields
behavior present in charged AdS black holes with and without global monopoles.

\section*{Acknowledgements}

DFA would like to thank Eloy Ay\'on--Beato for many enlightening conversations
and acknowledges support from CONACyT through Grant No. 404449.
This work was partially supported  by UNAM-DGAPA-PAPIIT, Grant No. 111617, and by the Ministry of Education and Science of RK, Grant No. 
BR05236322 and AP05133630.
We thank the anonymous referees for their critical and helpful comments.


\section*{References}
\begin{thebibliography}{99}

\bibitem{chem}
Kubiz\v{n}\'ak D, Mann R B and Teo M 2017
Black hole chemistry: thermodynamics with Lambda
{\it Class. Quantum Grav.} {\bf 34} 063001

\bibitem{cvj1}
Johnson C V 2014
Thermodynamic volumes for AdS-Taub--NUT and AdS-Taub-Bolt
{\it Class. Quantum Grav.} {\bf 31} 235003

\bibitem{smarr}
Smarr L 1973
Mass formula for Kerr black holes
{\it Phys. Rev. Lett.} {\bf 30} 71

\bibitem{hyun}
Hyun S, Jeong J, Park S-A and Yi S-H 2017
Thermodynamic volume and the extended Smarr relation
{\it JHEP} {\bf 04} 048

\bibitem{cvj2}
Johnson C V 2014
The extended thermodynamic phase structure of Taub--NUT and Taub-Bolt
{\it Class. Quantum Grav.} {\bf 31} 225005

\bibitem{surface}
Emparan R, Johnson C V and Myers R C 1999
Surface terms as counterterms in the AdS-CFT correspondence
{\it Phys. Rev.} D {\bf 60} 104001

\bibitem{Cmetrics}
Appels M., Gregory R. and Kubiz\v{n}\'ak D 2016
Thermodynamics of Accelerating Black Holes
{\it Phys. Rev. Lett.} {\bf 117} 131303

\bibitem{LTaubNUT}
Hennigar R A, Kubiz\v{n}\'ak D, Mann R B 2019
Thermodynamics of Lorentzian Taub--NUT spacetimes
{\it arXiv:1903.08668}

\bibitem{PS}
Pleba\'{n}ski J F and Schild A 1976
Complex relativity and double KS metrics
{\it Il Nuovo Cimento} B (1971-1996) {\bf 35} 35

\bibitem{condef}
Appels M., Gregory R. and Kubiz\v{n}\'ak D 2017
Black hole thermodynamics with conical defects
{\it JHEP} {\bf 05} 116

\bibitem{coshole}
Vilenkin A, Levin Y and Gruzinov A 2018
Cosmic strings and primordial black holes
{\it JCAP} {\bf 11} 008

\bibitem{vs}
Vilenkin A and Shellard EPS 1994 
{\it Cosmic Strings and Other Topological Defects}
(Cambridge: University Press)

\bibitem{skyrme}
Skyrme T H R 1961
Particle states of a quantized meson field
{\it Proc. R. Soc. London} A {\bf 262} 237

\bibitem{dhs1}
Droz S, Heusler M, and Straumann N 1991
New black hole solutions with hair
{\it Phys. Lett.} B {\bf 268} 371

\bibitem{dhs2}
Droz S, Heusler M, and Straumann N 1991
Stability analysis of self-gravitating skyrmions
{\it Phys. Lett.} B {\bf 271}, 61

\bibitem{cm}
Canfora F and Maeda H 2013
Hedgehog ansatz and its generalization for self-gravitating Skyrmions
{\it Phys. Rev.} D {\bf 87} 084049

\bibitem{bv}
Barriola M and Vilenkin A 1989
Gravitational field of a global monopole
{\it Phys. Rev. Lett.} {\bf 63} 341

\bibitem{global}
Gibbons G W 1991
{\it Proc. of the XII Autumn School of Physics (Lisbon)}
(Berlin: Springer) p 110; Self-gravitating magnetic monopoles, global monopoles and black holes {\it ArXiv:1109.3538}

\bibitem{lumps}
Kastor D and Traschen J 1992
Horizons inside classical lumps
{\it Phys. Rev.} D {\bf 46} 5399

\bibitem{hhadt}
Hawking S W and Horowitz G T 1996 
The gravitational Hamiltonian, action, entropy and surface terms
{\it Class. Quantum Grav.} {\bf 13} 1487

\bibitem{deng}
Deng G-M, Fan J, Li X and Huang Y-C 2018
Thermodynamics and phase transition of charged AdS black holes with a global monopole
{\it Int. J. Mod. Phys.} A {\bf 33} 1850022

\bibitem{ns}
Nucamendi U and Sudarsky D 1997
Quasi-asymptotically flat spacetimes and their ADM mass
{\it Class. Quantum Grav.} {\bf 14} 1309

\bibitem{heuslerbook}
Heusler M 1996 
{\it Black Hole uniqueness Theorems}
(Cambridge: University Press)

\bibitem{abbott}
Abbott L F and Deser S 1982
Stability of gravity with a cosmological constant
{\it Nucl. Phys.} B {\bf 195} 76

\bibitem{deser}
Abbott L F and Deser S 1982
Charge definition in non-Abelian gauge theories
{\it Phys. Lett.} B {\bf 116} 259

\bibitem{tekin}
Deser S and Tekin B 2002
Gravitational energy in quadratic curvature gravities
{\it Phys. Rev. Lett.} {\bf 89} 101101

\bibitem{KimKulkarniYi}
Kim W, Kulkarni S and Yi S-H 2013
Quasilocal Conserved Charges in a Covariant Theory of Gravity
{\it Phys. Rev. Lett.} {\bf 111} 081101

\bibitem{GimKimYi}
Gim Y, Kim W and Yi S-H 2014
The first law of thermodynamics in Lifshitz black holes revisited
{\it JHEP} {\bf 07} 002

\bibitem{hpads}
Hawking S W and Page D N 1983
Thermodynamics of black holes in anti-de Sitter space
{\it Commun. Math. Phys.} {\bf 87} 577

\bibitem{hs}
 Heusler M and Straumann N 1993
The first law of black hole physics for a class of nonlinear matter models
{\it Class. Quantum Grav.} {\bf 10} 1299

\bibitem{henneaux1}
 Henneaux M and Teitelboim C 1984
The Cosmological Constant as a Canonical Variable
{\it Phys. Lett.} B {\bf 143} 415

\bibitem{teitelboim}
 Teitelboim C 1985
The Cosmological Constant as a Thermodynamic Black Hole Parameter
{\it Phys. Lett.} B {\bf 158} 293

\bibitem{henneaux2}
 Henneaux M and Teitelboim C 1989
The Cosmological Constant and General Covariance
{\it Phys. Lett.} B {\bf 222} 195

\bibitem{miskovic-olea}
 Mi\v{s}kovi\'{c} O and Olea R 2009
Topological regularization and self-duality in four-dimensional anti-de Sitter gravity
{\it Phys. Rev.} D {\bf 79} 124020

\bibitem{kostas}
 de Haro S, Skenderis K and Solodukhin S N 2001
Holographic Reconstruction of Spacetime and Renormalization in the AdS/CFT Correspondence
{\it Commun. Math. Phys.} {\bf 217} 595

\bibitem{radu}
 Brihaye Y, Hartmann B and Radu E 2006
Global monopoles, cosmological constant and maximal mass conjecture
{\it Phys. Rev.} D {\bf 74} 025009

\bibitem{andrade}
 Andrade T and Ross S F 2013
Boundary conditions for scalars in Lifshitz
{\it Class. Quantum Grav.} {\bf 30} 065009

\bibitem{marika}
Taylor-Robinson M 2000
More on counterterms in the gravitational action and anomalies
{\it arXiv:hep-th/0002125}

\bibitem{kastor}
Kastor D, Ray S and Traschen J 2009
Enthalpy and the Mechanics of AdS Black Holes
{\it Class. Quantum Grav.} {\bf 26} 195011

\bibitem{jyw}
Jing J, Yu H and Wang Y 1993
Thermodynamics of a black hole with a global monopole
{\it Phys. Lett.} A {\bf 178} 59

\bibitem{yu}
Yu H-W 1994
Black hole thermodynamics and global monopoles
{\it Nucl. Phys.} B {\bf 430} 427

\bibitem{swallowtail}
Chamblin A, Emparan R, Johnson C V and Myers R C 1999
Charged AdS Black Holes and Catastrophic Holography
{\it Phys. Rev.} D {\bf 60} 064018

\bibitem{kubizmann}
Kubiz\v{n}\'ak D and Mann R B 2012
P-V criticality of charged AdS black holes
{\it JHEP} {\bf 07} 033

\end{thebibliography}
\end{document}